# Classification of Vehicles Based on Audio Signals using Quadratic Discriminant Analysis and High Energy Feature Vectors

*Ali Dalir [a], Ali Asghar Beheshti [b], Morteza Hosseini Masoom [c]*

[a] *Department of Electrical Engineering, Iran University of Science and Technology, Tehran, Iran. Email: Alidalir88@gmail.com*

[b] *Department of Electrical Engineering, Iran University of Science and Technology, Tehran, Iran. Email: abeheshti@iust.ac.ir*

[c] *Department of Mechanical Engineering, Babol Noshirvani University of Technology, Babol, Iran. Email: hosseini.mt@gmail.com*

*Corresponding Author: Alidalir88@gmail.com.*

*Abstract*

The focus of this paper is on classification of different vehicles using sound emanated from the vehicles. In this paper, quadratic discriminant analysis classifies audio signals of passing vehicles to bus, car, motor, and truck categories based on features such as short time energy, average zero cross rate, and pitch frequency of periodic segments of signals. Simulation results show that just by considering high energy feature vectors, better classification accuracy can be achieved due to the correspondence of low energy regions with noises of the background. To separate these elements, short time energy and average zero cross rate are used simultaneously. In our method, we have used a few features which are easy to be calculated in time domain and enable practical implementation of efficient classifier. Although, the computation complexity is low, the classification accuracy is comparable with other classification methods based on long feature vectors reported in literature for this problem.

## 1. Introduction

Vehicle identification while it is in motion is a prerequisite for traffic and speed management, classified vehicle count, traffic signal time optimization, gap/ headway measurement, and military purposes. Moving vehicles affect the environment in different ways. Vehicle emits heats, sounds, and magnetic field. There are many approaches investigated vehicle identification based on different kinds of signals. Image processing techniques are used to classify vehicles under the real time traffic management and for Intelligent Transportation Systems (ITS). Inductive loops based systems are widely used for determination of vehicle counts [1]. The most promising approach for vehicle identification is the one that is based on acoustic signals. Moving vehicles emit characteristic sounds. These sounds are generated from moving parts, frictions, winds, emissions, tires, etc. Assuming that similar vehicles which have the same working conditions generate the same sounds; then these sounds can be used to classify vehicles [2]. Vehicle classification based on sound signals have already attempted by researchers. Paulraj et al. (2013) [3] used autoregressive modeling algorithm for the analysis to extract the features from the recorded vehicle signals. Probabilistic neural network (PNN) models are developed to classify the vehicle type and its distance. Nooralahiyan et al. (1997) [4] used a directional microphone connected to a DAT (Digital Audio Tape) recorder. The digital signal was pre-processed by LPC (Linear Predictive Coding) parameter conversion based on autocorrelation analysis. A Time Delay Neural Network (TDNN) was chosen to classify individual travelling vehicles based on their speed independent acoustic signature to four broad categories: buses or Lorries, small or large saloons, various types of motorcycles, and light goods vehicles or vans. Ghiurcau and Rusu (2009) [5] presented a vehicle sound classification system using time encoded signal processing and recognition (TESPAR) method combined with the archetypes technique implemented on the Matlab platform. The experimental results show the efficiency of the TESPAR method when dealing with vehicle sounds. Effectiveness of features set is usually measured by how well they represent the signal; however, features should also satisfy several conditions: their number should be small (less than 15); also, to enable practical implementation they should be easy to calculate. There is a number of feature extraction methods used in vehicle classification. Some of them produce too many features for a single input vector like estimation of Power Spectrum Density [6], [7], [8], and some are too complicated like Principal Component Analysis [9], [7]. Nevertheless, no feature extraction method fulfills all our requirements, and there is no systematic way to select best feature according to our criterion. Harmonic line, Schur coefficients, and MEL filters, were presented in [10] each relating to properties of vehicle audio. Methods were compared in context of their separability and correct classification rate. Aljaafreh and Dong (2010) [2] investigated two feature extraction methods for acoustic signals from moving vehicles. The first one is based on spectrum distribution and the second one on wavelet packet transform. They evaluated the performance of different classifiers such as K-nearest neighbor algorithm (KNN) and support vector machine (SVM). It is found that for vehicle sound data, a discrete spectrum based feature extraction method outperforms wavelet packet transform method. Experimental results verified that support vector machine is an efficient classifier for vehicles. Other spectrum based feature extraction can be found in [11], [12], [13], [14], and [15]. Similarly, Discrete Wavelet Transform (DWT) is used in [16] and [17] to extract features using statistical parameters and energy content of the wavelet coefficients.

This paper describes an algorithm to classify audio signals of vehicles. Quadratic discriminant analysis uses feature vectors of periodic segments with elements: short time energy, average zero cross rate, and pitch frequency to distinguish between vehicles' signals. If we separate feature vectors with high energy from others properly and train classifier with these vectors, better classification accuracy will be achieved due to the correspondence of low energy regions and noises of the background, as simulation results confirm. Obviously high energy feature vectors have less zero cross rate. Considering this point is the basic idea for separation criterion. Our innovation in separation is assuming vectors by energy larger than $E[E_n]$, average of all segments' energy, and ZCR less than $E[Z_n]$, average of all segments' ZCR, are high energy elements. We have used three features which are easy to be calculated in time domain, which enable practical implementation of efficient classifier. The cost paid is lower, but still satisfactory, classification accuracy than other classification applications reported in literature for this problem. Sections of paper are as follow: After introduction first, we determine important parameters and features. Then we study quadratic discriminant analysis and high energy element separation criterion in section III. Simulation results show improvement in classification accuracy for separation criterion. Finally, the paper ends by conclusion.

## 2. Feature Extraction

Changing the characteristics of an audio signal over time seems normal. For example, Changes in the signal's peak value or rate of sign changes of amplitude are such a time varying features. Although these changes exist in signals, short time analysis is a method to illustrate them as obvious way. Basic assumption of this method is based on these changes occur slowly; then we divide signal into smaller parts and features are extracted.

## 2.1. Short Time Energy

This feature indicates how the signal amplitude changes over the time which is defined as follows:

$$E_n = \sum_{m=-\infty}^{\infty}[x(m)w(n-m)]^2 \quad (1)$$

Where $x(n)$ is the audio signal and $w(n)$ is the window that slides along the audio sequence selecting the interval to be involved in the computation.

## 2.2. Short Time Average Zero Cross Rate

This feature indicates how the signal sign changes over the time which is defined as follows:

$$Z_n = \sum_{m=-\infty}^{\infty}|sgn[x(m)] - sgn[x(m-1)]| \, w(n-m) \quad (2)$$

Where

$$w(n) = 1/2 \quad \quad 0 \leq n \leq N-1$$

$$= 0 \quad \quad otherwise \quad \quad (3)$$

The rate at which signal sign changes occur is a simple measure of the frequency content of a signal. This is particularly true for narrowband signals. For example, a sinusoidal signal of frequency $F_0$, sampled at a rate $F_s$, has $2\,F_0/F_s$ average rate of zero-crossings. Thus, the average ZCR gives a resonable way to estimate the frequency of a sine wave. Audio signals are broadband signals and the interpretation of average ZCR is therefore much less precise. However, rough estimates of spectral properties can be obtained using a represantation based on the short time energy and zero cross rate together. This representation was proposed by Reddy, and studied by Vicens [18] as the basis for a large-scale speech recognition system.

## 2.3. Pitch Frequency

Pitch period estimation (or fundamental frequency) is one of the most important problems in speech processing. Pitch detectors are used in vocoders [19], speaker identification and verification systems [20, 21]. Because of its importance, many solutions to this problem have been proposed [22]. In this paper we use the fact that the autocorrelation function for periodic segments attains a maximum at samples 0, $\pm T$, $\pm 2T$, … where $T$ is pitch period. Since, there may be exist other autocorrelation peaks except those due to the periodicity, the simple procedure of picking the largest peak at sample $T$ in the autocorrelation will fail to estimate Pitch period. To avoid this problem it is useful to process the signal so as to make periodicity more prominent while suppressing other distracting features of the signal. Spectrum flattener with the objective to bring each harmonic to the same amplitude level as in the case of a periodic impulse train has been used. Specifically, center clipped signal has been used in computing autocorrelation function. In the scheme proposed, similar to [23], the center clipped signal is obtained by following nonlinear transformation

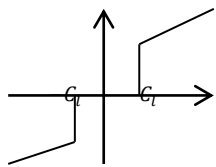

Fig. 1: center clipping function

For samples above $C_l$, the output is equal to the input minus the clipping level. For samples below the clipping level the output is zero. We set clipping level, $C_l$, equal to 68% of the minimum of two maximum amplitudes found in both the first third and last third of the audio segment; denoted respectively by $max_1$ and $max_2$. However, for the un-periodic segment, there are no strong autocorrelation periodicity peaks and it seems to be a high frequency noise-like waveform [22]. We assume segments with weak peaks below 30 % clipped signal's energy, $E_{clip_n} = \sum_{m=-\infty}^{\infty}[x_{clip}(m)w(n-m)]^2$, are un-periodic. Finally pitch frequencies of frames smoothed through median operation. Following figures present features.

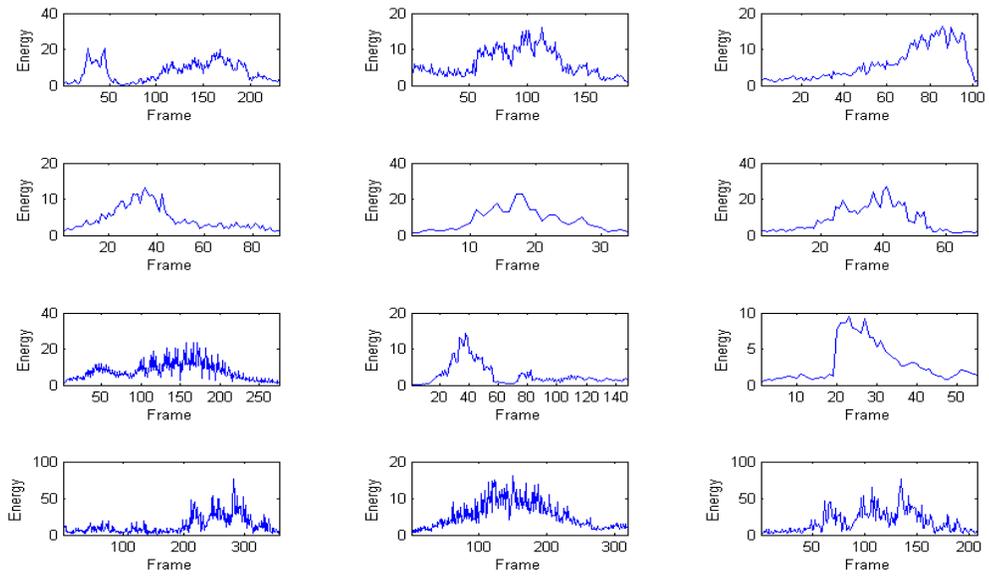

Fig. 2: Energy for Bus in first row for three different sounds, for Car in the second row, for Motor in the third row, and for Truck in the fourth row.

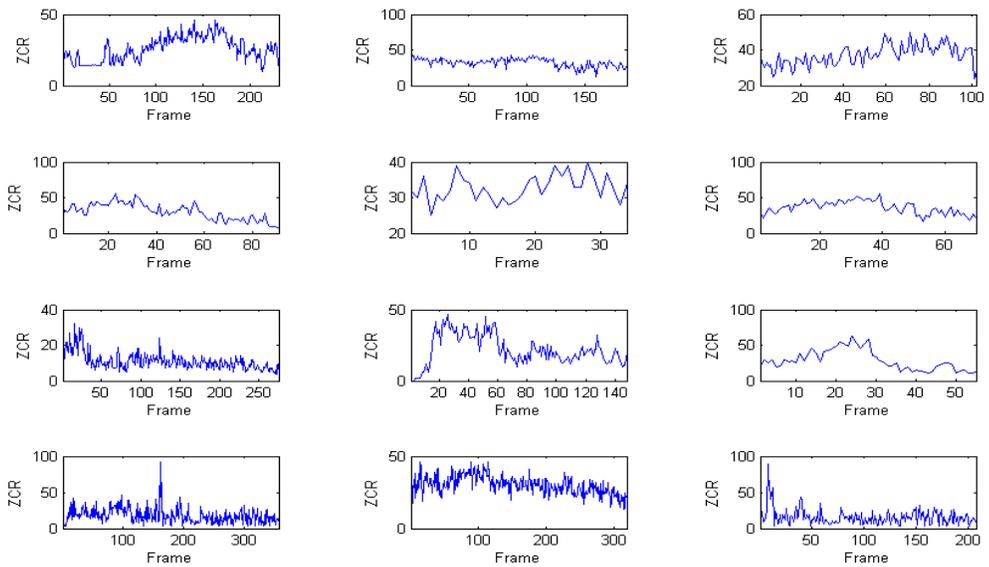

Fig. 3: ZCR for Bus in first row for three different sounds, for Car in the second row, for Motor in the third row, and for Truck in the fourth row.

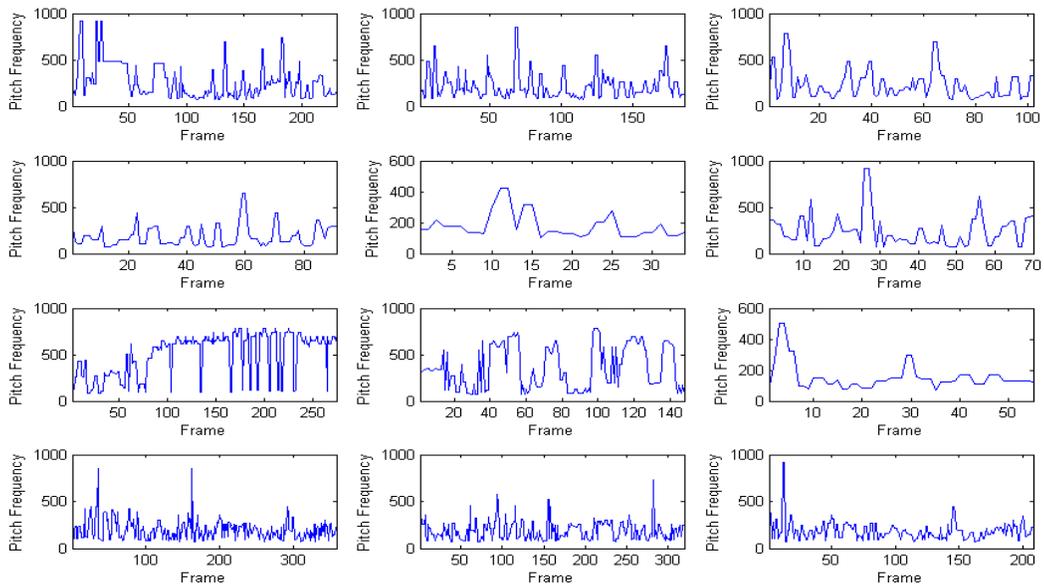

Fig. 4: Pitch Frequency for Bus in first row for three different sounds, for Car in the second row, for Motor in the third row, and for Truck in the fourth row.

Features are extracted from rectangular windowed segment of length 15 Milliseconds and overlapping of 5 Milliseconds between consecutive windows. Feature vectors of periodic segments are used for training and classification. After determining the important features and parameters for audio signal modeling, we will review quadratic discriminant analysis and high energy separation criterion will be described.

## 3. Classifier and Separation Criterion

Now the discriminant boundaries between classes should be determined. Quadratic discriminant analysis is used to discriminate the classes. The classifier that separate $K$ classes has $K$ quadratic functions as follow [24]:

$$y_k(X) = X^T Q X + V^T X + v_0 \qquad (4)$$

The coefficient $v_0$ is constant coefficient, $V$ is vector of linear coefficients, and $Q$ is matrix containing quadratic coefficients. Feature vector $X$ will belong to class $k$, if $y_k(X) > y_j(X); \forall j \neq k$. In classification step, after feature vectors were extracted, we are able to define class of each vector separately. Finally, signal will allocate to a class which achieved great supply in feature vector classification phase. In other words, probability of belonging to each class is proportional to abundance of that class in allocated labels to vectors by quadratic discriminant analysis.

### 3.1. High Energy Vector Separation Criterion

Up to now all periodic vectors are considered in training and classification. However, low energy regions corresponding with noises of the background consist of high frequency variations may not be suited for vehicle classification. Hence by ignoring mentioned vectors, an improvement in classification accuracy can be achieved. On the other hand, it can be observed that when amplitude is high zero cross rate decreases. So, high energy elements have low zero cross rate, and low energy vectors have fast sign alternation. A criterion for separation elements with high energy and low zero cross rate can be defined as follow:

$$E_n > \alpha . E[E_n] \text{ And } Z_n < \zeta . E[Z_n] \qquad (5)$$

Where $E[E_n]$ and $E[Z_n]$ denote average of all segments' energy, ZCR. The coefficient α is more than 1, and ζ is less than 1. The more α and the less ζ than unit, the less data coincides in condition. In this paper α = ζ = 1 are chosen. Figure 5 presents a block diagram of our method.

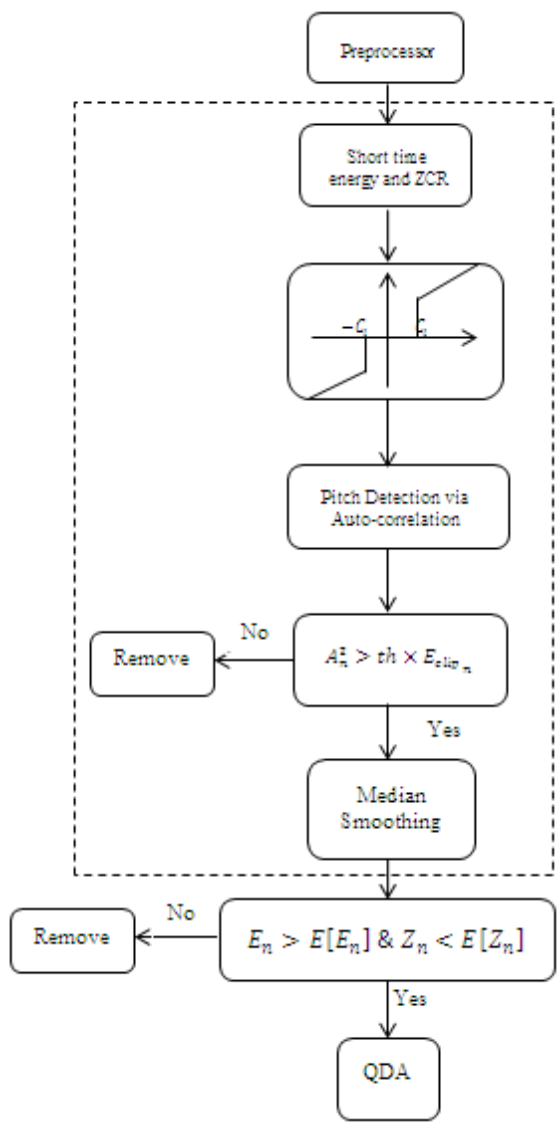

Fig. 5: Block diagram of the proposed scheme for vehicle classification

## 4. Results and Discussion

Simultaneous arrival of different types vehicles, overtaking at the study location, and random back ground noises were the different challenges encountered in the data collection. Sound emitted by vehicles was captured for a divided road carrying moderate traffic. Roadside recording was performed on clear sunny working days. The recordings' sites were selected to capture the sound of moving vehicles without much acceleration change. Environmental back ground noise is at moderate level. The conditions under which data set is recorded are realistic. Our algorithm is implemented on the Matlab 7.10.0 (R2010a) platform.

*Preprocessing:* Each signal is normalized such that $-1 \leq x(m) \leq 1$. Windowed signal segments are filtered by $4^{th}$ degree Butterworth low pass filter.

The data set includes four classes; bus, car, motor, and truck. All signals sampled at 11025 rate. Each signal is segmented by rectangular window of length 15 milliseconds and overlapping of 5 milliseconds between consecutive windows. Vectors containing short time energy, average zero cross rate, and pitch frequency are calculated for each segment and those with 0 pitch frequencies are removed. We used *k*-Fold cross validation with *k*=10 in our applications. In this method data is randomly divided into a test and training set *k* different times [25] and the average error across all *k* trials is computed. Table 1 provides the values of some important parameters employed in this work. Classification accuracies for following classifier have been provided in table 2.

- least square classifier,
- *k*-nearest neighbor classifier,
- quadratic and linear discriminant analysis when training and testing have been done just based on vectors which be able to satisfy the (5) conditions; denoted respectively by $QDA^{**}$ and $LDA^{**}$,

- SVM classifier with harmonic line based feature vectors [10],
- SVM classifier with Schur filter coefficients based feature vectors [10],
- SVM classifier with Mel filter coefficients based feature vectors [10].

We see improvement in classification accuracy has been achieved for quadratic discriminant analysis and separation criterion, $QDA^{**}$. The correct classification rate depends not only on the feature extraction method, but also on the type of the classifier. This is the reason that why not we see a considerable improvement for $LDA^{**}$.

Table 3 shows computation complexity in terms of feature space dimensionality for our proposed algorithm and last 3 methods in table 2. In comparison, we should note that the accuracy of $QDA^{**}$ with 3 short time parameters based feature vectors is comparable with SVM with 16 Schur coefficients. SVM with 12 Mel filter coefficients accuracy is 80% and accuracy of $QDA^{**}$ is 68.75%. Although the computation complexity of $QDA^{**}$ is low, but the classification accuracy is lower than SVM with Mel filter coefficients.

| | |
|---|---|
| Signal sample rate | 11025 sample/sec |
| Length of window for classification | 165 sample |
| Overlap between windows for classification | 55 sample |
| Type of window | Rectangular |
| Preprocessor | $4^{th}$ degree Butterworth LPF |
| LPF's cut off frequency | 4 $KHz$ |
| Clipping level | $0.68 * \min\{max_1, max_2\}$ |
| Maximum pitch frequency | $f <= 1000\ Hz$ |

| Periodic/Un-periodic threshold | $0.3 * E_{clip_n}$ |
|---|---|
| Number of frames over which median operates | 3 |
| α and ζ | 1 |
| Number of cross fold validation | 10 |
| examples per class: [bus, car, motor, truck] | [50, 70, 80, 40] |

Table 1: Values of some important parameters

| Classifier | Accuracy (%) |
|---|---|
| Least Square | 30 |
| kNN, $k$=25, Cosine | $52.09 \pm 2.2$ |
| kNN, $k$=25, Euclidian | $53.33 \pm 3.83$ |
| LDA | 46.66 |
| $LDA^{**}$ | 49.58 |
| QDA | 56.25 |
| $QDA^{**}$ | 68.75 |
| SVM with harmonic line | 83 |
| SVM with Schur coefficients | 68 |
| SVM with Mel filter coefficients | 80 |

Table 2: Classification accuracy

| Feature extraction method | Number of features |
|---|---|
| Short time analysis | 3 |
| harmonic lines | 5 |
| Schur coefficients | 16 |
| Mel filter coefficients | 12 |

Table 3: Number of features

Additionally confusion matrix, contain information about actual and predicted classifications done by $QDA^{**}$ classification system, shown in table 4.

| Actual | Predicted | | | |
|---|---|---|---|---|
| | Bus vehicles | Car vehicles | Motor vehicles | Truck vehicles |
| Bus vehicles | 27 | 20 | 3 | 0 |
| Car vehicles | 0 | 70 | 0 | 0 |
| Motor vehicles | 16 | 12 | 48 | 4 |
| Truck vehicles | 9 | 11 | 0 | 20 |

Table 4: Representation of confusion matrix

Although recording conditions were realistic, but we should examine proposed algorithm when environmental noises increase like rainy days. Since we use high energy feature vectors, $QDA^{**}$ can save its effectiveness.

**Conclusion**

In the research reported in this paper, QDA with a criterion to separate high energy feature vectors was applied on the task of classification of vehicles based on audio signals and the simplest method with satisfactory accuracy was evaluated.

Short time energy, average zero cross rate, and pitch frequency made feature vectors; then, signals were classified using quadratic discriminant analysis. We found that classification accuracy improved from 56.25% to 68.75% just by considering components which have (5) conditions due to the correspondence of low energy regions and noises of the background. At last, the obtained accuracy was 68.75% by using three features while the 80% correct classification rate was achieved by SVM with 12 Mel coefficients [10]. The results strongly suggest that proposed method can aid the practical implementation.

## References

[1] J. George, and et al. '' Exploring Sound Signature for Vehicle Detection and Classification Using ANN'' International Journal on Soft Computing (IJSC) Vol.4, No.2, May 2013.

[2] A. Aljaafreh, and L. Dong ''an Evaluation of Feature Extraction Methods for Vehicle Classification Based on Acoustic Signals'' International Conference on Networking, Sensing and Control (ICNSC), 2010.

[3] M. P. Paulraj, and et al. ''Moving Vehicle Recognition and Classification Based on Time Domain Approach'' Procedia Engineering, Volume 53, 2013, Pages 405–410.

[4] Y. Nooralahiyan, and et al. ''Field Trial of Acoustic Signature Analysis for Vehicle Classification'' Transportation Research Part C: Emerging Technologies, Volume 5, Issues 3–4, August–October 1997, Pages 165–177.

[5] M. V. Ghiurcau, C. Rusu, ''Vehicle Sound Classification Application and Low Pass Filtering Influence'' In proceeding of International Symposium on Signals, Circuits and Systems, 2009, ISSCS 2009.

[6] M. Wellman, N. Srour, and D. Hillis, "Feature Extraction and Fusion of Acoustic and Seismic Sensors for Target Identification," in Society of Photo-Optical Instrumentation Engineers (SPIE) Conference Series, ser. Society of Photo-Optical Instrumentation Engineers (SPIE) Conference Series, G. Yonas, Ed, vol. 3081. SPIE, 1997, pp. 139–145.

[7] M. Wellman, "Acoustic Feature Extraction for a Neural Network Classifier." DTIC Document, Tech. Rep., 1997.

[8] M. F. Duarte and Y. H. Hu, "Vehicle Classification in Distributed Sensor Networks," Journal of Parallel and Distributed Computing, vol. 64, pp. 826–838, 2004.

[9] H. Wu, M. Siegel, and P. Khosla, "Vehicle Sound Signature Recognition by Frequency Vector Principal Component Analysis," in Instrumentation and Measurement Technology Conference, 1998. IMTC/98. Conference Proceedings. IEEE, vol. 1, May 1998, pp. 429 –434 vol.1.

[10] M. Gorski, J. Zarzycki, ''Feature Extraction in Vehicle Classification'' International Conference on Signals and Electronic Systems (ICSES), 2012.

[11] H. Wu, M. Siegel, and P. Khosla, "Distributed classification of acoustic targets wireless audio-sensor networks," Computer Networks, vol. 52, no. 13, pp. 2582–2593, Sep. 2008.

[12] ——, "Vehicle classification in distributed sensor networks," Journal of Parallel and Distributed Computing, vol. 64, no. 7, pp. 826–838, July 2004.

[13] Y. Seung S., K. Yoon G., and H. Choi, "Distributed and efficient classifiers for wireless audio-sensor networks," in 5th International Conference on Volume, Apr. 2008.

[14] S. S. Yang, Y. G. Kim1, and H. Choi, "Vehicle identification using discrete spectrums in wireless sensor networks," Journal of Networks, vol. 3, no. 4, pp. 51–63, Apr. 2008.

[15] H. Wu, M. Siegel, and P. Khosla, "Vehicle sound signature recognition by frequency vector principal component analysis," IEEE Trans. Instrum. Meas., vol. 48, no. 5, pp. 1005–1009, Oct. 1999.

[16] C. H. C. K. R. E. G. G. R. and M. T. J, "Wavelet-based ground vehicle recognition using acoustic signals," Journal of Parallel and Distributed Computing, vol. 2762, no. 434, pp. 434–445, 1996.

[17] A. H. Khandoker, D. T. H. Lai, R. K. Begg, and M. Palaniswami,"Wavelet-based feature extraction for support vector machines for screening balance impairments in the elderly," vol. 15, no. 4, pp. 587–597, 2007.

[18] P. J. Vicens, "Aspects of Speech Recognition by Computer," Ph.D. Thesis, Stanford Univ., AI Memo No. 85, Comp. Sci. Dept., Stanford Univ., 1969.

[19] J. L. Flanagan, *Speech Analysis, Synthesis and Perception*, 2$^{nd}$ Ed., Springer Verlag, N.Y., 1972.

[20] B. S. Atal, "Automatic Speaker Recognition Based on Pitch Contours," *J. Acoust. Soc. Am.,* Vol. 52, pp. 1687-1697, December 1972.

[21] A. E. Rosenberg and M. R. Sambur, "New Techniques for Automatic Speaker Verification," *IEEE Trans. Acoust, Speech, and Signal Proc.,* Vol. 23, pp. 169-176, April 1975.

[22] L. R. Rabiner, R. W. Schafer, Digital Processing of Speech Signals. Englewood Cliffs, N.J., Prentice Hall.

[23] M. M. Sondhi, "New Methods of Pitch Extraction," IEEE Trans. Audio and Electroacoustics, Vol. 16, No. 2, pp. 262-266, June 1968.

[24] K. Fukunaga, Introduction to Statistical Pattern Recognition. San Diego, Academic Press, 1990, pp. 153-154.

[25] R. Kohavi, F. Provost, Glossary of terms, Editorial for the Special Issue on Applications of Machine Learning and the Knowledge Discovery Process, vol. 30, No. 2–3, 1998.